\documentclass[12pt]{article}
\usepackage{epsf,epsfig,rotating}

\begin{document}


\def\jrn#1#2#3#4{{#1} {\bf #2} (#4) #3}

\def\PRL{Phys.Rev.Lett.}
\def\PLB{Phys.Lett.B}
\def\PRD{Phys.Rev.D}
\def\NPB{Nucl.Phys.B}
\def\YaF{Yad.Fiz.}

\begin{titlepage}

\title{
          Using the $e^{\pm}\mu^{\mp}~+~E^{miss}_T$ and $l^+l^- ~+~E^{miss}_T$ 
          Signatures in the Search for Supersymmetry and 
          Constraining the MSSM model at LHC
}

\author{  Yu.Andreev, N.Krasnikov and A.Toropin \\
          Institute for Nuclear Research RAS,  Moscow, 117312, Russia
}

\maketitle
  
\begin{abstract}

We study the $e^{\pm}\mu^{\mp} ~+~E^{miss}_T$ and $l^+l^- ~+~E^{miss}_T$
signatures $(l = e, ~\mu)$ for different values of $\tan\beta$ in the 
mSUGRA model. With $\tan\beta$ rising, we observe a characteristic 
change in the shape of dilepton mass spectra in $l^+l^- + E^{miss}_T $ 
versus $e^{\pm}\mu^{\mp}$ final states reflecting the decrease of 
$\tilde{\chi}^0_2 \rightarrow l^+l^- \tilde{\chi^0_1}$ branching ratio. 
We also study the non mSUGRA modifications of the CMS test point 
LM1 with arbitrary relations among gaugino and higgsino masses. For such 
modifications of the mSUGRA test point LM1 the number of lepton events 
depends rather strongly on the relations among gaugino and higgsino 
masses and in some modifications of the test point LM1 the signatures with 
leptons and $E_{T}^{miss}$ do not lead to the SUSY discovery and the single 
SUSY discovery signature remains the signature with 
$n \geq 2~jets~ + ~E^{miss}_T~ +~ no~ leptons$.

\end{abstract}
   
\end{titlepage}

\section{
\label{Intr}
                        Introduction
}

One of the goals of the Large Hadron Collider (LHC) \cite{1} is the discovery 
of supersymmetry (SUSY). The squark and gluino decays produce missing transverse 
energy $E^{miss}_T$ from lightest stable superparticle (LSP) plus multiple jets 
and isolated leptons \cite{1}. One of the most interesting and widely discussed
signatures for SUSY discovery at the LHC is the signature with two opposite 
charge and the same flavour leptons \cite{Baer}: $l^+l^- ~+ ~E^{miss}_T$. 
The main reason of such interest is that neutralino decays into leptons and LSP 
$\tilde{\chi}^0_2 \rightarrow l^+l^- \tilde{\chi}^0_1$ contribute to this 
signature and the distribution of the $l^+l^-$ invariant mass $m_{inv}(l^+l^-)$ 
has the edge structure \cite{Abdu} that allows to determine some combination of 
the SUSY masses. 

The signature $e^{\pm}\mu^{\mp} ~+ ~E^{miss}_T$ can be realized when $\chi^0_2$ 
decays into $\tau$ pair which is only relevant at large $\tan\beta$ \cite{Daniel}.
Also as it is shown in Ref.\cite{Abdu} at the level of CMSJET \cite{cmsjet} 
simulation the use of $e^{\pm}\mu^{\mp}~+~E^{miss}_T$ signature for large 
$\tan\beta$ allows to obtain nontrivial information on parameters of the decay 
$\tilde{\chi}^0_2 \rightarrow \tilde{\tau}\tau \rightarrow \tau\tau\tilde{\chi}^0_1  
\rightarrow e^{\pm}\mu^{\mp}\tilde{\chi}^0_1\nu\nu\bar{\nu}\bar{\nu}$. 
On the other hand, the $e^{\pm}\mu^{\mp} ~+ ~E^{miss}_T$ (with an arbitrary number 
of jets) can be used for the detection of lepton flavour violation 
in slepton decays \cite{Kras1994}, \cite{Depp} at the LHC.

In the Minimal Supersymmetric Model (MSSM) \cite{Msugra} supersymmetry 
 is broken at some high scale $M$ by generic 
soft terms, so in general all soft SUSY breaking terms are arbitrary 
which complicates the analysis and spoils the predictive power of the theory. 
In the Minimal Supergravity Model (mSUGRA) \cite{Msugra} 
the universality of the different soft parameters 
at the Grand Unified Theory (GUT) scale 
$M_{GUT} \approx 2 \cdot 10^{16}$ ~GeV is postulated. 
Namely, all the spin zero 
particle masses (squarks, sleptons, higgses) are postulated to be equal to 
the universal value $m_0$ at the GUT scale. All gaugino particle masses 
are postulated to be equal to the universal value $m_{1/2}$ at GUT scale. 
Also the coefficients in front of quadratic and cubic SUSY soft breaking  
terms are postulated to be equal. The renormalization group equations are 
used to relate GUT and electroweak scales. The equations for the 
determination of a nontrivial minimum of the electroweak potential are 
used to decrease the number of the unknown parameters by two. 
So the mSUGRA model depends on five unknown parameters. At present,
the more or 
less standard choice of free parameters in the mSUGRA model includes 
$m_0, m_{1/2}, \tan \beta , A$ and $ sign(\mu) $ \cite{Msugra}. 
All sparticle masses depend on these parameters.

In Ref.\cite{ABKT} the possibility to detect SUSY and 
lepton flavour violation
using the $e^{\pm}\mu^{\mp} + E^{miss}_T$ signature at the  
LHC for the Compact Muon Solenoid (CMS) detector at the level of 
full detector simulation was studied. Note also Ref.\cite{Tricomi}  
where the SUSY mSUGRA CMS detector discovery potential was 
investigated for the 
$l^+l^- ~+~ E^{miss}_T$ signature at the level of full detector simulation.

In this paper we   study the  $e^{\pm}\mu^{\mp} ~+~E^{miss}_T$ and 
$l^+l^- ~+~E^{miss}_T$ signatures $(l = e, ~\mu)$ for different values of 
$\tan\beta $ in the mSUGRA model. With $\tan \beta$ rising, we observe a 
characteristic change in the shape of dilepton mass 
spectra in $l^+l^- + E^{miss}_T $ versus $e^{\pm}\mu^{\mp}$ 
final states reflecting the decrease of 
$\tilde{\chi}^0_2 \rightarrow l^+l^- \tilde{\chi^0_1}$ 
branching ratio. We also study the non mSUGRA modifications of the CMS 
test point LM1 with arbitrary relations among gaugino 
and higgsino masses. For such modifications of the 
mSUGRA test point the number of lepton 
events depends rather strongly on the relations among gaugino and  
higgsino masses and in some modifications of  test point LM1
the signatures with leptons and $E_{T}^{miss}$ do not lead to the SUSY 
discovery and the single SUSY discovery signature remains
the signature with $n \geq 2~jets~ + ~E^{miss}_T~ +~ no~ leptons$.

The organization of the paper is the following. Section \ref{Tede}  describes 
some useful technical details of performed simulations. In Section \ref{Basi}
the backgrounds and cuts used to suppress the backgrounds are discussed. 
Section \ref{Resca} contains  the results of numerical calculations concerning 
the possibility to detect SUSY and constrain the SUSY parameters 
using the $e^{\pm}\mu^{\mp}~+~E_T^{miss}$ and $l^+l^- ~+~ E^{miss}_T$
signatures. Section \ref{Ressim} contains the results of the simulations for 
different values of $\tan \beta$. In Section \ref{Ressim2} we discuss the 
results of the simulation of the MSSM with nonuniversal gaugino and higgsino 
masses. Section \ref{Conc} contains concluding remarks.

\section{
\label{Tede}
                        Simulation details
}

The coupling constants and cross sections in the leading order (LO) 
approximation for SUSY processes and backgrounds were calculated with 
ISASUGRA 7.69 \cite{ISASUGRA}, PYTHIA 6.227 \cite{Pythia} and 
CompHEP 4.2pl \cite{compHEP}. For the calculation of the 
next-to-leading order (NLO) corrections to the SUSY cross sections 
the PROSPINO \cite{PROSPINO} code was used. For considered signal events 
and backgrounds the NLO corrections are known and 
the values of NLO cross sections (or $k$-factors) were used
for normalization of the numerical results. 
We used the full simulation results of Ref.\cite{ABKT} for the estimation of 
the number of background events. The CMS fast simulation code 
$FAMOS\_1\_4\_0$ \cite{CMSSOFT} was used for the estimation of signal events.

The reconstructed electrons and muons were passed through packages defining 
lepton isolation criteria. For each electron and muon the following 
parameters  were  defined:
\begin{itemize}
\item{$TrackIsolation$ is a number of additional tracks with $p_T >$ 2~GeV/$c$ 
  inside a cone with 
$R \equiv \sqrt{\Delta \eta^{2} + \Delta \Phi^{2}} < $ 0.3} around the lepton.
\item{$CaloIsolation$ is a ratio of energy deposited
 in  the calorimeters (electromagnetic (ECAL) + hadronic (HCAL)) inside a cone 
with $ R = 0.13 $ around given track to the energy 
deposited inside a cone with $ R = 0.3 $.}
\item{$HEratio$ is defined as a ratio of energy deposited 
in the HCAL inside a cone with $R = 0.13$ 
to the energy deposited in the ECAL inside the same cone.} 
\item{$EPratio$ is a  ratio of 
energy deposited in the ECAL inside a cone 
with $R = 0.13$  to the momentum of the reconstructed track.}
\end{itemize}

\section{
\label{Basi}
                   Signal selection and backgrounds 
}

The SUSY production $ pp \rightarrow \tilde{q}\tilde{q}^{'}, \tilde{g}
\tilde{g}, \tilde{q}\tilde{g}$ with subsequent decays
\begin{equation}
\tilde{q} \rightarrow q^{'}\tilde{\chi}^{\pm}_{1,2}  \\,
\end{equation}
\begin{equation}
\tilde{g} \rightarrow q \bar{q}^{'} \tilde{\chi}^{\pm}_{1,2}\\,
\end{equation}
\begin{equation}
\tilde{\chi}^{+}_{1,2} \rightarrow \tilde{\chi}^0_1 e^+(\mu^{-})\nu \\,
\end{equation}
\begin{equation}
\tilde{\chi}^{-}_{1,2} \rightarrow \tilde{\chi}^0_1\mu^{-}(e^{-})\nu \\,
\end{equation}
lead to the event topologies $e^{\pm}\mu^{\mp} ~+ ~E^{miss}_T$ and 
$l^+l^- ~+~E^{miss}_T$.  Note that 
in the MSSM with lepton flavour conservation neutralino decays into leptons 
$\tilde{\chi}^0_{2,3,4} \rightarrow l^+l^- \tilde{\chi}^0_1$ 
($l \equiv e,\mu$) do not 
contribute into the $e^{\pm}\mu^{\mp} ~+~E^{miss}_T$  signature. 
 The main backgrounds which contribute to the $e^{\pm}\mu^{\mp}$ events 
are: ${\rm t \bar t}$, WW, WZ, ZZ, Wt, Z${\rm b \bar b}$, $\tau \bar{\tau}$ 
and Z+jet. 
It is found that ${\rm t \bar t}$ is the largest background and it
gives more than 50\% contribution to the total background. 
In this paper we used the results of Ref.\cite{ABKT} for the estimation of the 
number of background events. 

In the analysis the events with the following isolation criteria 
for electrons were used:
$TrackIsolation < 1.0$, $CaloIsolation > 0.85$, $0.85 < EPratio < 2.0$, 
$HEratio <0.25$.
The same criteria for muons were the following: 
$TrackIsolation < 1.0$, $CaloIsolation > 0.50$, $EPratio < 0.20$, 
$HEratio > 0.70$. 
These numbers were adjusted by studying electron and muon tracks  in 
the process $ pp \rightarrow WW \rightarrow 2l$.

The selection cuts are the following:
 
\begin{itemize}
\item { 
cut on leptons: $p_T^{lept} > p_T^{lept, 0}$, $|\eta | < 2.4$, 
lepton isolation within ${\Delta}R < 0.3$ cone}
\end{itemize}

\begin{itemize}
\item { cut on missing transverse energy: $E_T^{miss} > E_T^{miss, 0}$. }
\end{itemize}

Here $p_T^{lept, 0}$ and $E_T^{miss, 0}$ are corresponding thresholds.

\section{
\label{Resca}
                        Use of the $e^{\pm}\mu^{\mp}~+~E^{miss}_T$
                        signature for the SUSY detection
}

In this section we remind the main results of Ref.\cite{ABKT} 
on SUSY detection using the $e^{\pm}\mu^{\mp} + E^{miss}_T$ signature. 
The possibility to detect SUSY using the
CMS test points LM1 - LM9 \cite{LMPoints}  chosen for the detailed 
study of SUSY detection at the CMS was investigated in Ref.\cite{ABKT}. 
This study was based on the counting the expected
number of events for both the Standard Model (SM) and the mSUGRA model.
The parameters of the CMS test points LM1 - LM9 are given in Table \ref{Points}. 

\begin{table}[!htb]
\begin{center}
\caption{ The parameters of the CMS test points.}
\begin{tabular}{cccccc}
\hline
\hline
~Point~~~& $ m_0$ (GeV)  & $ m_{1/2}$(GeV) & $\tan{\beta}$ & $sign(\mu)$ & 
$A_0 $    \\
\hline
LM1 & 60 & 250 & 10 & + & 0 \\
LM2 & 185 & 350 & 35 & + & 0  \\
LM3 & 330 & 240 & 20 & + & 0 \\
LM4 & 210 & 285 & 10 & + & 0  \\
LM5 & 230 & 360 & 10 & + & 0  \\
LM6 & 85 & 400 & 10 & + & 0 \\
LM7 & 3000 & 230 & 10 & + & 0  \\
LM8 & 500 & 300 & 10 & + & -300 \\
LM9 & 1450 & 175 & 50 & + & 0 \\
\hline
\hline
\end{tabular}
\label{Points}
\end{center}
\end{table}

For the point LM1 (the point LM1 coincides with the post-WMAP point B 
\cite{Batt}) it was found that the cuts with $p_T^{lept} >$ 20~GeV/$c$, 
$E_T^{miss} >$ 300~GeV are close to the optimal ones (the highest significance 
with the best signal/background ratio). The 
results for the luminosity ${\cal L} = ~10~{fb}^{-1}$  
are presented in Table \ref{Bgsgemu}. For other CMS SUSY test points 
LM2 - LM9 the results with the same cuts are presented in Table \ref{Fv_res}. 
The significances  in this table are the following:
$S_{c12}=2(\sqrt{N_S+N_B}-\sqrt{N_B})$ \cite{Bity} and 
$S_{cL}~=~$ $\sqrt{2((N_S~+~N_B)~ln(1~+~N_S/N_B)~-~N_S)}$ \cite{BartV}.

\begin{table}[!htb]
\begin{center}
\caption{ The expected number of events for backgrounds and for 
signal at the point LM1, ${\cal L} = 10~fb^{-1}$, 
$e^{\pm}\mu^{\mp}~+~E^{miss}_T$ signature.}
\begin{tabular}{lcc}
\hline
\hline
Process & 2 isolated leptons, $p_T^{lept} >$ 20~GeV/$c$ & $E^{miss}_T >$ 300~GeV \\
\hline
${\rm t \bar t}$ & 39679 & 79  \\
\hline
WW & 4356 & 4  \\
\hline
WZ   & 334 & 2 \\
\hline
ZZ & 38 &  0   \\
\hline
Wt & 3823 & 2  \\
\hline
Z${\rm b \bar b}$ & 315 & 0  \\
\hline
Z+jet & 1082 & 6  \\
\hline
DY2$\tau$ & 7564 & 0  \\
\hline
SM background & 57191 & 93  \\
\hline
LM1 Signal & 1054 & 329  \\
\hline
\hline
\end{tabular}
\label{Bgsgemu}
\end{center}
\end{table}

\begin{table}[!htb]
\begin{center}
\caption{ The number of signal events and significances for the cuts 
with $p_T^{lept}~>$~20~GeV/$c$  and $E_{T}^{miss}~>$~300~GeV
for ${\cal L} = 10~fb^{-1}$, signature $e^{\pm}\mu^{\mp}~+~E^{miss}_T$.
The number of the SM background events $N_{B}$ = 93 (see Table \ref{Bgsgemu}).
}
\begin{tabular}{crrr}
\hline
\hline
~~Point~~ & ~$N$ events & ~~~$S_{c12}$ & ~~~$S_{cL}$ \\
\hline
LM1 & 329 & 21.8 & 24.9  \\
LM2 &  94 &  8.1 &  8.6  \\
LM3 & 402 & 25.2 & 29.2  \\
LM4 & 301 & 20.4 & 23.1  \\
LM5 &  91 &  7.8 &  8.3  \\
LM6 & 222 & 16.2 & 18.0  \\
LM7 &  14 &  1.4 &  1.4  \\
LM8 & 234 & 16.9 & 18.8  \\
LM9 & 137 & 11.0 & 11.9  \\
\hline
\hline
\end{tabular}
\label{Fv_res}
\end{center}
\end{table}

It was found from the Tables \ref{Bgsgemu}-\ref{Fv_res} that for the point LM1
the significances are $S_{c12} = 21.8$ and
$S_{cL} = 24.9$ for the $e^{\pm}\mu^{\mp}~+~E^{miss}_T$ signature.

\begin{figure}[hbtp]
  \begin{center}
    \resizebox{12cm}{!}{\includegraphics{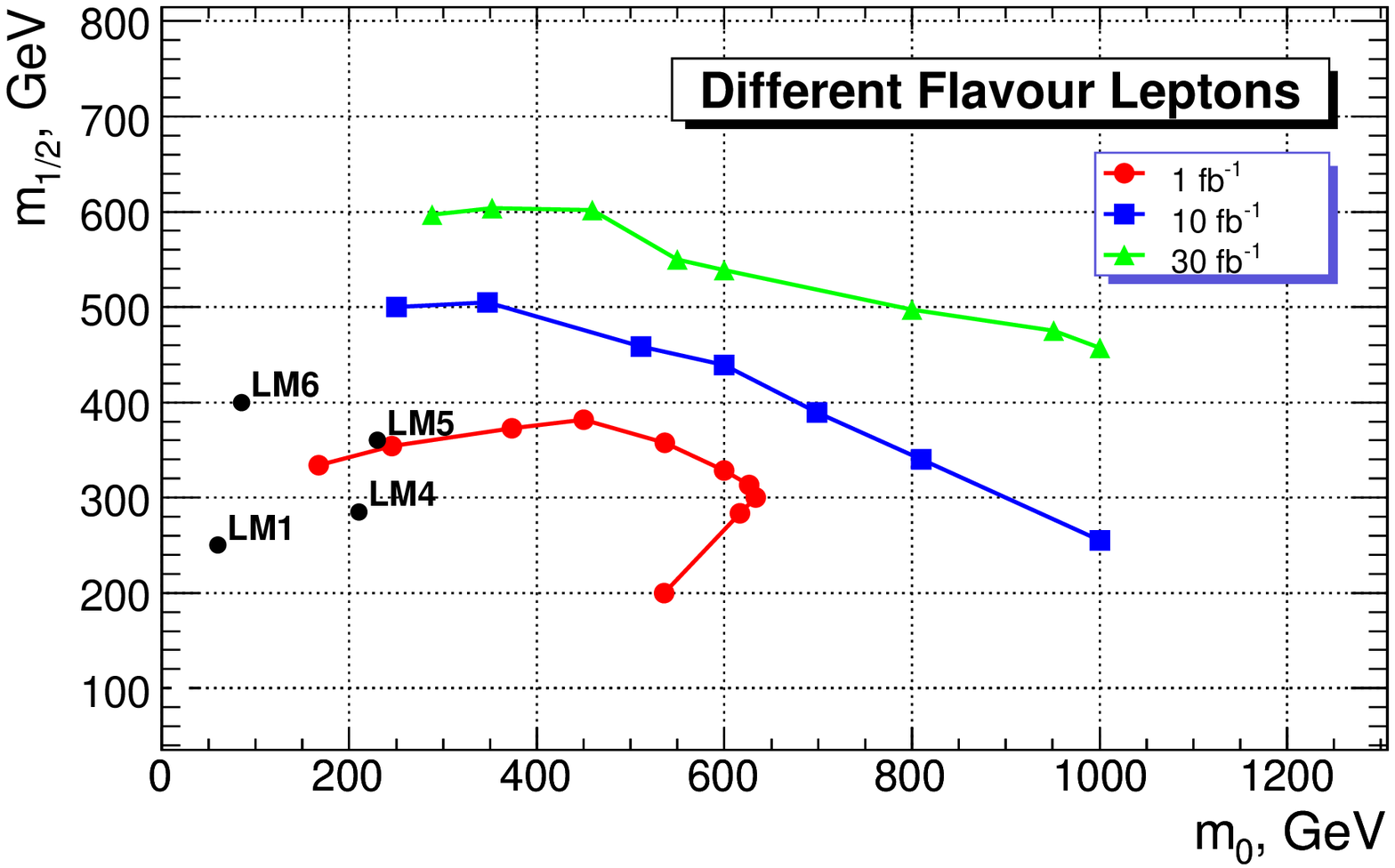}}
    \caption{ $e^+\mu^{-} ~+ ~e^{-}\mu^{+}$ discovery plot 
      for $\tan{\beta} = 10$, $sign (\mu) = +$, $ A = 0$.
      Two isolated leptons with  $p_T^{lept} >$ 20~GeV/$c$ and 
      $E^{miss}_T >$ 300~GeV are selected. }
    \label{fv_emDP} 
  \end{center}
 \end{figure}

The supersymmetry discovery potential for the mSUGRA model with 
$\tan\beta = 10$, $sign(\mu) = +$ ~in the $(m_0, ~m_{1/2})$ plane 
(generalization of the point LM1) using the CMS fast simulation program 
$FAMOS\_1\_4\_0$ \cite{CMSSOFT} was also studied.
The CMS discovery potential contours for ${\cal L} = 1, 10$ and $30~fb^{-1}$ 
for the signature $e^{\pm}\mu^{\mp}~+~E^{miss}_T$ are shown in Fig.\ref{fv_emDP}.

\section{
\label{Ressim} 
                        The shape of the dilepton mass distribution
                        in the mSUGRA parameter space
}

At the LHC neutralinos $\tilde{\chi}^0_2$ are dominantly produced in the 
decay chain of gluino and squarks, for instance 
$\tilde{g} \rightarrow q\bar{q} \tilde{\chi}^0_2$ 
or $\tilde{q} \rightarrow q \tilde{\chi}^0_2$  and 
$\tilde{\chi}^0_j \rightarrow \tilde{\chi}^0_2Z^0$.
Within the mSUGRA model in the domain of relatively small 
$m_0~ (m_0 \leq 0.6 m_{1/2})$ the 
following leptonic decays of $\tilde{\chi}^0_2$ are the most important:
\begin{equation}
\tilde{\chi}^0_2  \rightarrow \tilde{\chi}^0_1 l^+l^- \\, 
\end{equation} 
\begin{equation}
\tilde{\chi}^0_2  \rightarrow \tilde{l}_L l \\, 
\end{equation} 
\begin{equation}
\tilde{\chi}^0_2  \rightarrow \tilde{l}_R l \\, 
\end{equation} 
\begin{equation}
\tilde{\chi}^0_2  \rightarrow \tilde{\tau}_1 \tau \\, 
\end{equation} 
\begin{equation}
\tilde{\chi}^0_2  \rightarrow \tilde{\chi}^0_1 \tilde{\tau}_{1} \tau_{1}. 
\end{equation} 

\begin{figure}[hbtp]
  \begin{center}
    \resizebox{12cm}{!}{\includegraphics{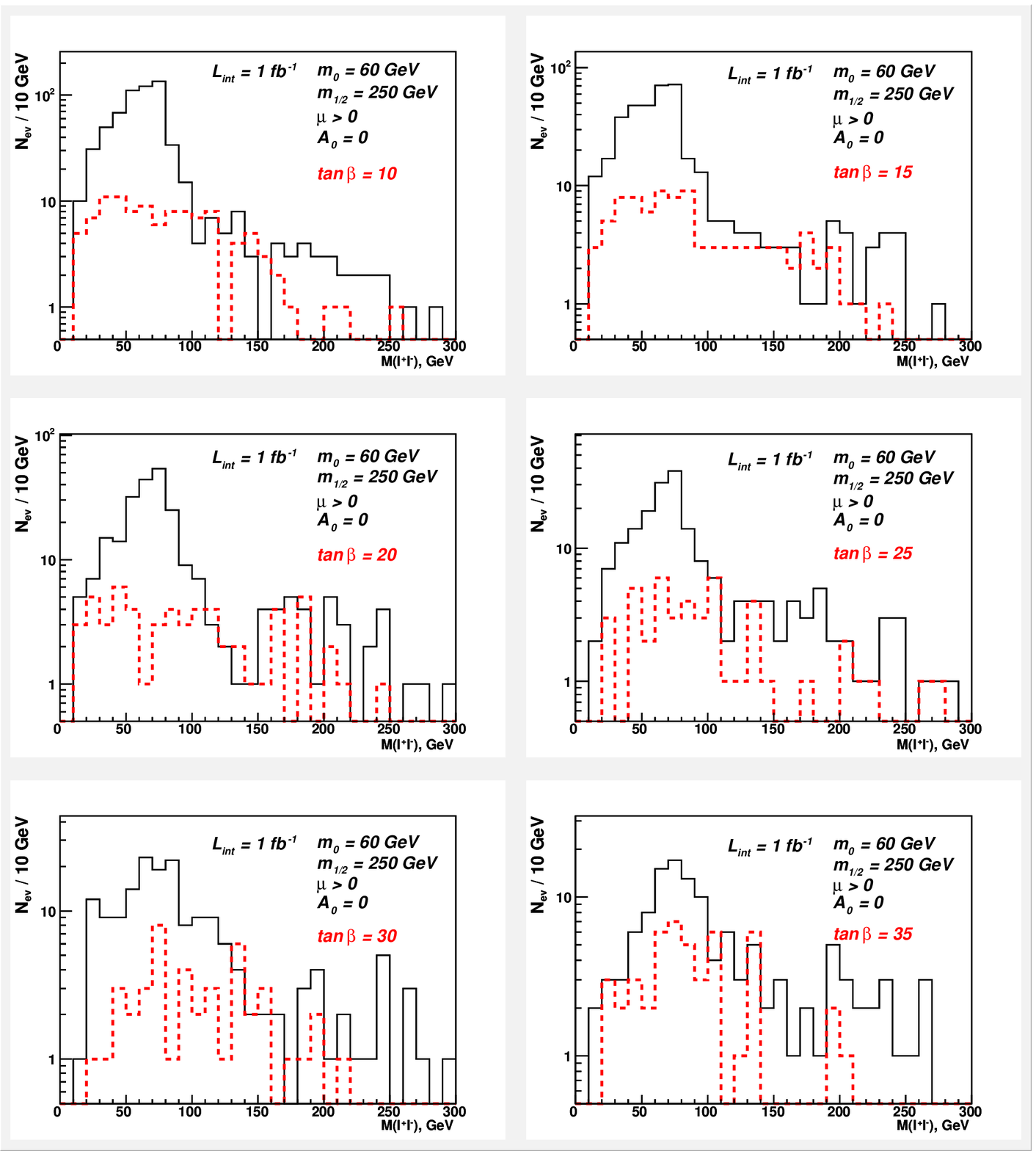}}
    \caption{ The invariant mass distributions of $l^+l^-$   
      (solid line) and $e^{\pm}\mu^{\mp}$ (dotted line) lepton pairs 
      at the mSUGRA point $m_0 = 60~GeV, M_{1/2} = 250~GeV, ~\mu > 0, 
      A = 0$ with various $\tan\beta = 10, ~15, ~20, ~25, ~30, ~35$. }
    \label{pLM1_MinvBSM6} 
  \end{center}
\end{figure}

In mSUGRA for the case when $\tilde{\chi}^0_2$ decays into lepton and slepton
are kinematically allowed, the sleptons decay directly into LSP 
with  $Br(\tilde{l}_{L,R} \rightarrow l\tilde{\chi}^0_1)\sim  1$ 
and $Br(\tilde{\tau}_1 \rightarrow \tau \tilde{\chi}^0_1) 
\sim 1$. With the increase of $\tan{\beta}$ the $\tilde{\tau}_1$ mass is 
decreased and as a consequence the branching ratio   
$Br( \tilde{\chi}^0_2 \rightarrow \tilde{\tau}_1 \tau)$ is increased whereas 
the branching ratio 
$Br(\tilde{\chi}^0_2 \rightarrow \tilde{l}_{R,L} l)$ is decreased, 
see Fig.\ref{pLM1_MinvBSM6}. 

The process 
$\tilde{\chi}^0_2 \rightarrow \tilde{l}l \rightarrow ll\tilde{\chi}^0_1$ 
has the edge structure for the distribution of lepton-pair invariant 
mass $m_{ll}$ and the edge mass $m^{max}_{inv}(l^+l^-)$ is expressed by 
slepton mass $m_{\tilde{l}}$ and neutralino masses 
$m_{\tilde{\chi}^0_{1,2}}$ as follows \cite{Abdu}:
\begin{equation}
(m^{max}_{inv}(l^+l^-))^2 = m^2_{\tilde{\chi}^0_2}(1 - \frac{m^2_{\tilde{l}}}
{m^2_{\tilde{\chi}^0_2  }})(1 - \frac{m^2_{\tilde{\chi}^0_1}}
{m^2_{\tilde{l}}  }   )
\end{equation}
The corresponding  $m^{max}_{inv}(\tau^+\tau^-)$ edge maximum due to 
$\tilde{\chi}^0_2$ decays to stau's has a maximum at: 
\begin{equation}
 (m^{max}_{inv}(\tau^+\tau^-))^2 = m^2_{\tilde{\chi}^0_2}(1 - \frac{m^2_{\tilde{\tau_1}}}
 {m^2_{\tilde{\chi}^0_2  }})(1 - \frac{m^2_{\tilde{\chi}^0_1}}
 {m^2_{\tilde{\tau_1}}  }   )
 \end{equation}
But the spectrum of the dilepton same flavour opposite sign channel 
proceeding through $\tau \rightarrow l \nu \bar{\nu}$ decays 
is not so pronounced as the spectrum from 
$ \tilde{\chi}^0_2 \rightarrow \tilde{l}l \rightarrow ll\tilde{\chi}^0_1$ 
decay due to the missing momemtum taken by four neutrinos from $\tau$ decays.
In particular the $l^+l^{-}$ invariant mass is distributed in the lower mass 
region below the expected ditau kinematical point. Thus, a distinctive 
feature of the $m_{inv}(l^+l^{'-})$ spectrum from 
$\tilde{\chi}^0_2$ decays to staus is distributed in the lower mass 
region below the expected ditau kinematical end point.
With increase of $\tan\beta$, due to significant increase of $\tilde{\chi}^0_2$ 
decays into stau's and the corresponding decrease of decays into selectrons 
and smuons, a deterioration of the sharpeness of the $l^+l^-$ dilepton edge 
takes place \cite{Daniel}. Another consequence of the change in the relative 
branching ratios of these decays is that the event rate difference between 
$e^+e^- ~+~ \mu^+\mu^-$ and $e^+\mu^- ~+~ \mu^+e^-$ channels decreases 
\cite{Daniel}. To illustrate the behaviour of the $m_{inv}(l^+l^-)$ and 
$m_{inv}(l^+l^{'-})$ spectra at different $\tan\beta$ we investigated several 
mSUGRA points, namely:

a. The generalization of the point LM1 ($m_0 = 60~GeV, m_{1/2} = 250~GeV$,
$A =0, sign(\mu) = +,  \tan\beta = 10$) with 
$\tan\beta = 15, ~20, ~25, ~30,~35$.

b. The points with $m_0 = 100~GeV, m_{1/2} = 300~GeV,A = 300~GeV$, 
$sign(\mu) = +$ and $\tan\beta = 10, ~15, ~20, ~25, ~30, ~35$.

c. The points with $m_0 = 300~GeV, m_{1/2} = 300~GeV,~A = 0$, 
$sign(\mu) = +$ and $\tan\beta = 10, ~15, ~20, ~25, ~30, ~35$.

d. The points with $m_0 = 2500~GeV, m_{1/2} = 250~GeV,A = 0$, 
$sign(\mu) = +$ and $\tan\beta = 10, ~15, ~20, ~25, ~30, ~35$.

For the points (a) and (b) sleptons are lighter than $\tilde{\chi}^0_2$ 
that  leads to the existence of well-defined edge 
structure in the $m_{inv}(l^+l^-)$ structure for not very large values of 
$\tan\beta$. For the points (c) and (d) the sleptons are heavier 
than $\tilde{\chi}^0_2$ and the dominant decay mode is $\tilde \chi^0_2 
\rightarrow Z ~+~\tilde{\chi}^0_1$ with the branching ratio close to 100\%. 

\begin{figure}[hbtp]
  \begin{center}
    \resizebox{12cm}{!}{\includegraphics{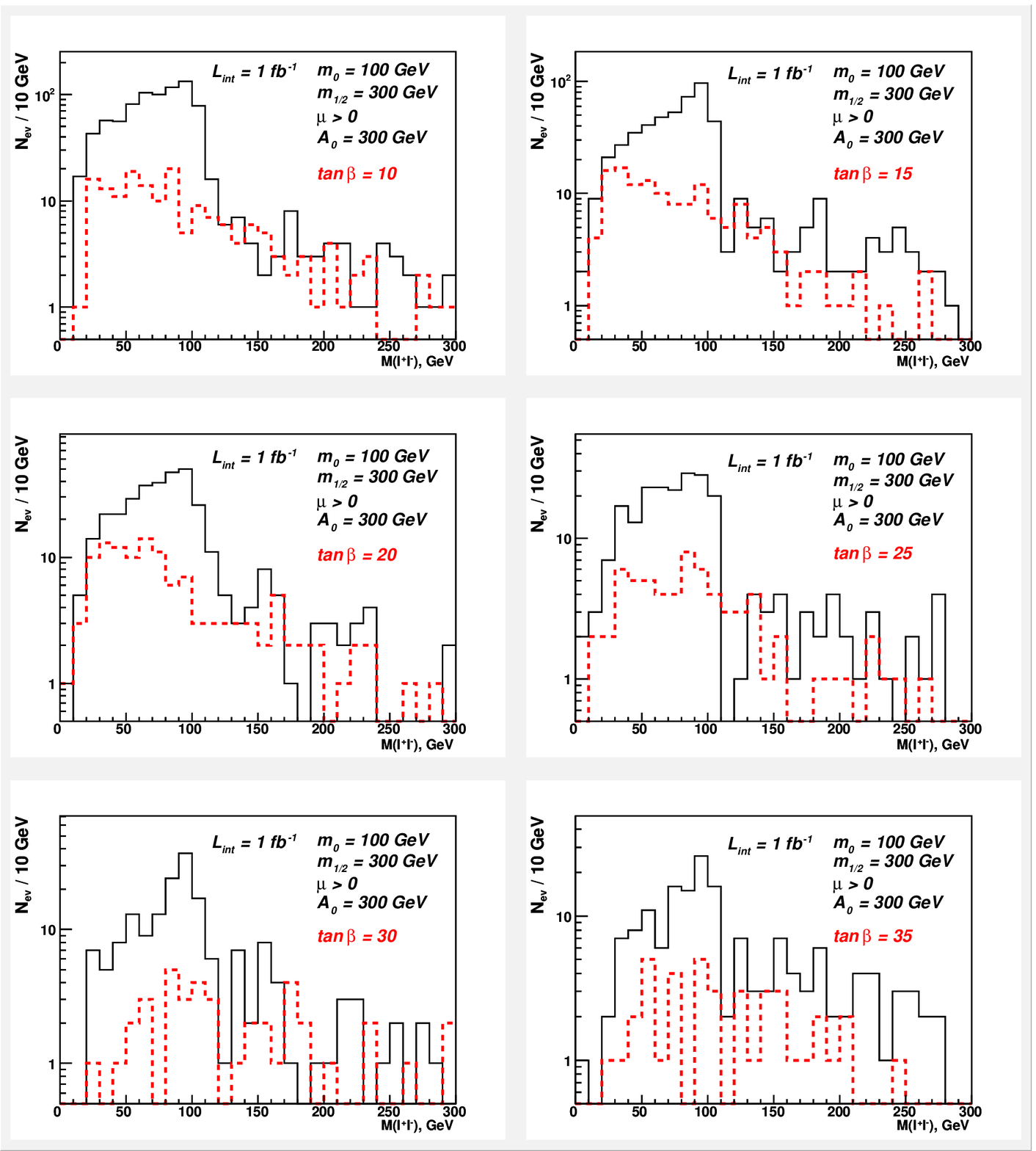}}
    \caption{ The invariant mass distributions of $l^+l^-$  
      (solid line) and $e^{\pm}\mu^{\mp}$ (dotted line) lepton pairs 
      at the mSUGRA point $m_0 = 100~GeV, M_{1/2} = 300~GeV, ~\mu > 0, 
      A = 300~GeV$ with various $\tan\beta = 10, ~15, ~20, ~25, ~30, ~35$. }
    \label{pATL_MinvBSM6} 
  \end{center}
\end{figure}

\begin{figure}[hbtp]
  \begin{center}
    \resizebox{12cm}{!}{\includegraphics{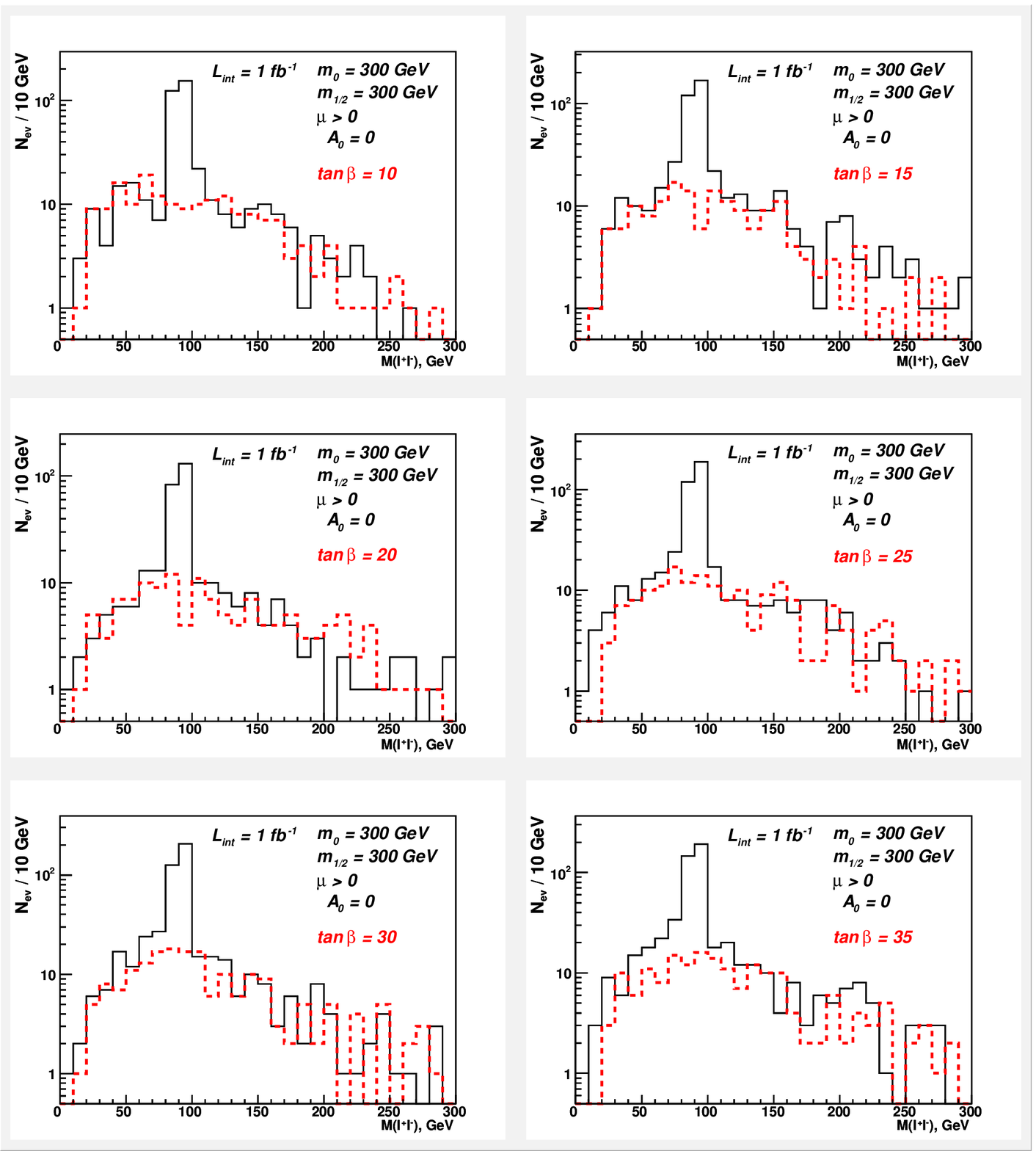}}
    \caption{ The invariant mass distributions of $l^+l^-$   
      (solid line) and $e^{\pm}\mu^{\mp}$ (dotted line) lepton pairs 
      at the mSUGRA point $m_0 = 300~GeV, M_{1/2} = 300~GeV, ~\mu > 0, 
      A = 0$ with various $\tan\beta = 10, ~15, ~20, ~25, ~30, ~35$. }
      \label{p300_MinvBSM6} 
  \end{center}
\end{figure}

\begin{figure}[hbtp]
  \begin{center}
    \resizebox{12cm}{!}{\includegraphics{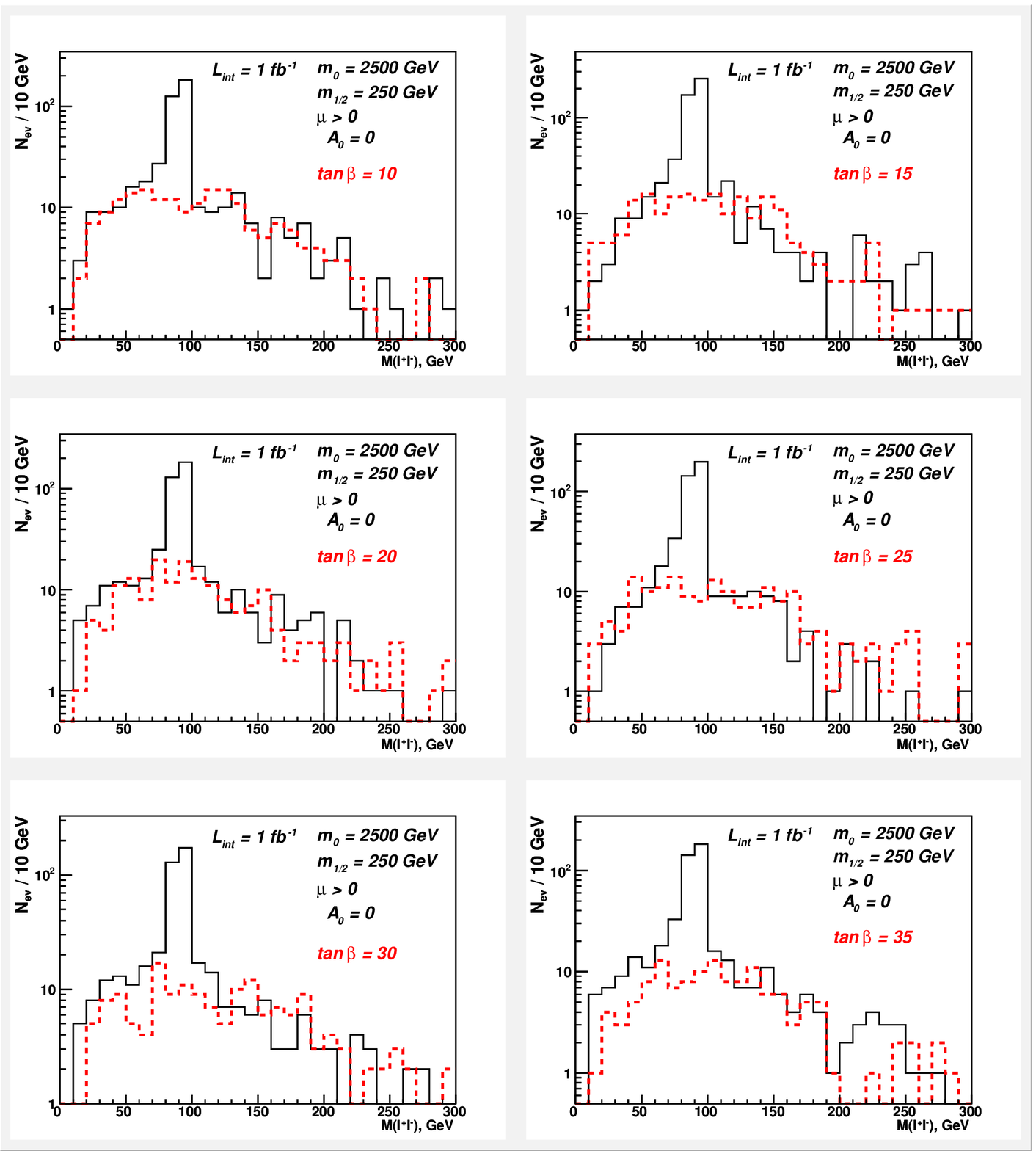}}
    \caption{ The invariant mass distributions of $l^+l^-$   
      (solid line) and $e^{\pm}\mu^{\mp}$ (dotted line) lepton pairs 
      at the mSUGRA point $m_0 = 2500~GeV, M_{1/2} = 250~GeV, ~\mu > 0, 
      A = 0$ with various $\tan\beta = 10, ~15, ~20, ~25, ~30, ~35$. }
    \label{p2500_MinvBSM6} 
  \end{center}
\end{figure}

\begin{figure}[hbtp]
  \begin{center}
    \resizebox{12cm}{!}{\includegraphics{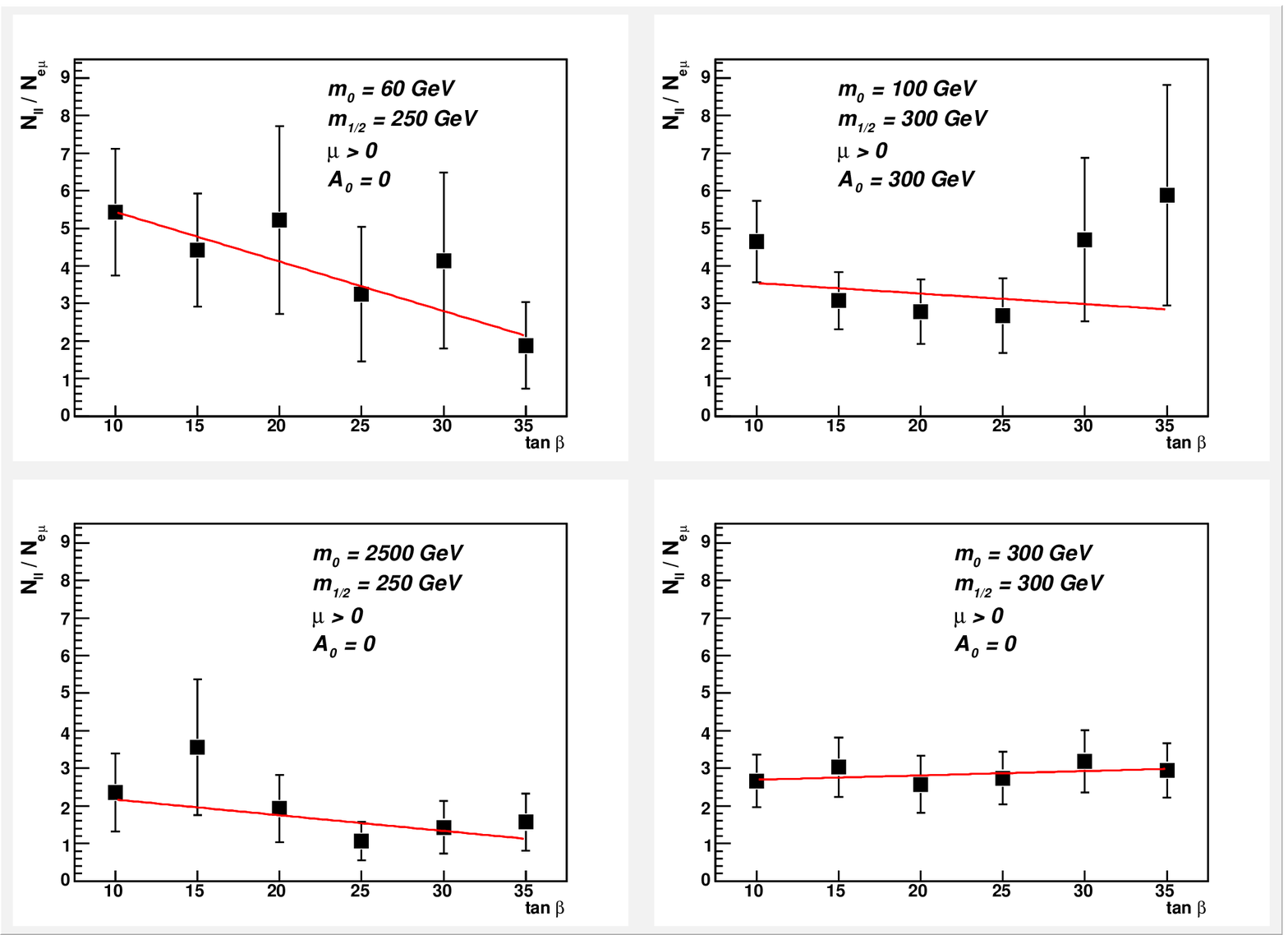}}
    \caption{ The dependence of the ratio $N_S(l^+l^-)/N_S(e^{\pm}\mu^{\mp})~$ 
      ($N_S( )$ is the number of events after the cuts
      $p_T^{lept} > 20~GeV, ~E^{miss}_T > ~300~GeV$) on 
      $\tan\beta $ for different scenarios. Here the error is due to 
      finite number of simulated events. }
    \label{pRatio4} 
  \end{center}
\end{figure}

\begin{figure}[hbtp]
  \begin{center}
    \resizebox{12cm}{!}{\includegraphics{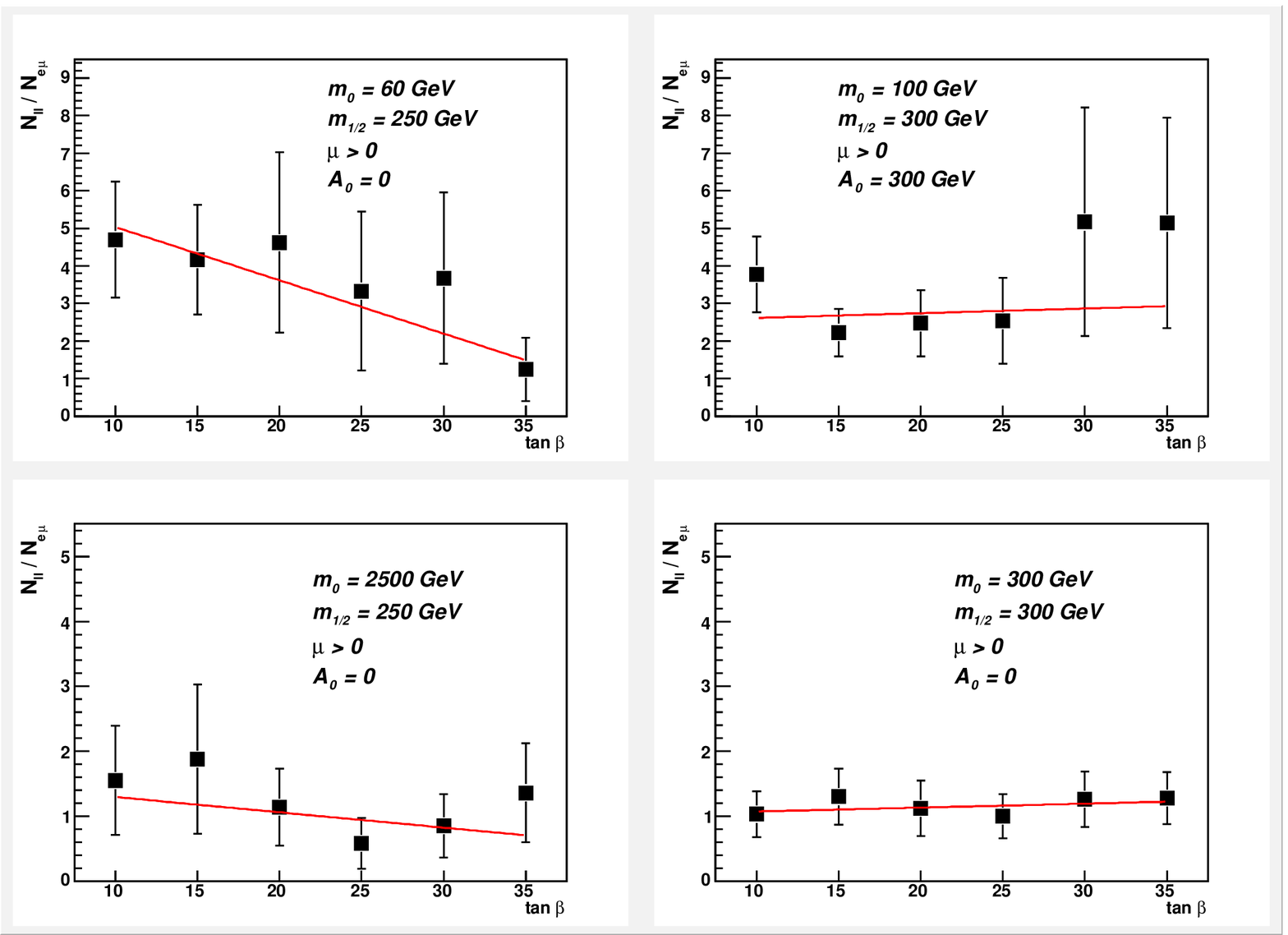}}
    \caption{ The dependence of the ratio $N_S(l^+l^-)/N_S(e^{\pm}\mu^{\mp})~$ 
      ($N_S( )$ is the number of events after the cuts 
      $p_T^{lept} > 20~GeV, ~E^{miss}_T > ~300~GeV, 
      |M_{inv} - M_Z| > 15~GeV $), on $\tan\beta$ for different scenarios.
      Here the error is due to finite number of simulated events. }
    \label{pRatioMz4} 
  \end{center}
\end{figure}

For these mSUGRA points we have made calculations using FAMOS fast simulation 
program for the integrated luminosity ${\cal L} = 1~fb^{-1}$. 
The results of our simulations are presented in 
Figs.\ref{pLM1_MinvBSM6} - \ref{pRatioMz4}.  

As one can see from Figs.\ref{pLM1_MinvBSM6} - \ref{pRatioMz4} 
with $\tan\beta$ rising there is a 
characteristic change in the shape of dilepton $l^+l^-$ mass spectra versus 
$e^{\pm}\mu^{\mp}$ mass spectra resulting in the increase of the ratio of 
$N_S(l^+l^-)/N_S(e^{\pm}\mu^{\mp})$ signal events.  
As it has been mentioned before this behaviour is due to the increase of
$Br(\tilde{\chi}^0_2 \rightarrow \tilde{\tau}\tau) $ and decrease of 
$Br(\tilde{\chi}^0_2 \rightarrow \tilde{l} l)$  with the rising of $\tan\beta$.

\section{
\label{Ressim2} 
                 The dileptons for the case of nonuniversal gaugino masses 
}

In this section we present the results of our simulations for the case of 
nonuniversal gaugino masses. Namely, we consider the deformations of the test 
point LM1 violating gaugino mass universality assumption, 
see Table \ref{Factors}.

\begin{table}[!htb]
\begin{center}
\caption{ Values of factors for masses increasing  (unity if not indicated) 
in comparison with the masses of the point LM1.}
\begin{tabular}{ccccccc}
\hline
\hline
~Point~~~ & $ M_1$ & $ M_2$ & $M_3$ & $\mu $ & $m_2$ & $ m_{\tilde{q}}$ \\
\hline
1  &  & 2 & &   &   &  \\
2  &  & 4 & &   &   &  \\
3  &  & 8 & &   &   &  \\
4 &  & 2 & & 2  &   &  \\
5  &  & 4 & & 4  &   &  \\
6 &  & 8 & & 8 &   &  \\
7 & 2 & 4 & 5 & 4 &   &  \\
8 & 3 & 4 & 5 & 4  &   &  \\
9 & 2 & 4 &5 & 4  & 5  &  \\
10 & 3  & 4 &5 & 4  & 5  &  \\
11 &  &  & &   &   & 10 \\
12 &2  & 2 & &   &   & 10 \\
\hline
\hline
\end{tabular}
\label{Factors}
\end{center}
\end{table}

In Table \ref{Factors}  $M_1, ~M_2$ and $M_3$ are 
$U(1), ~SU(2)$ and $SU(3)$ 
gaugino masses correspondingly and $\mu$ is higgsino mass parameter. 
$m_2$ means masses of the first two squark and slepton generations and 
$m_{\tilde{q}}$ means all squark and slepton masses. 

\begin{figure}[hbtp]
  \begin{center}
    \resizebox{13cm}{!}{\includegraphics{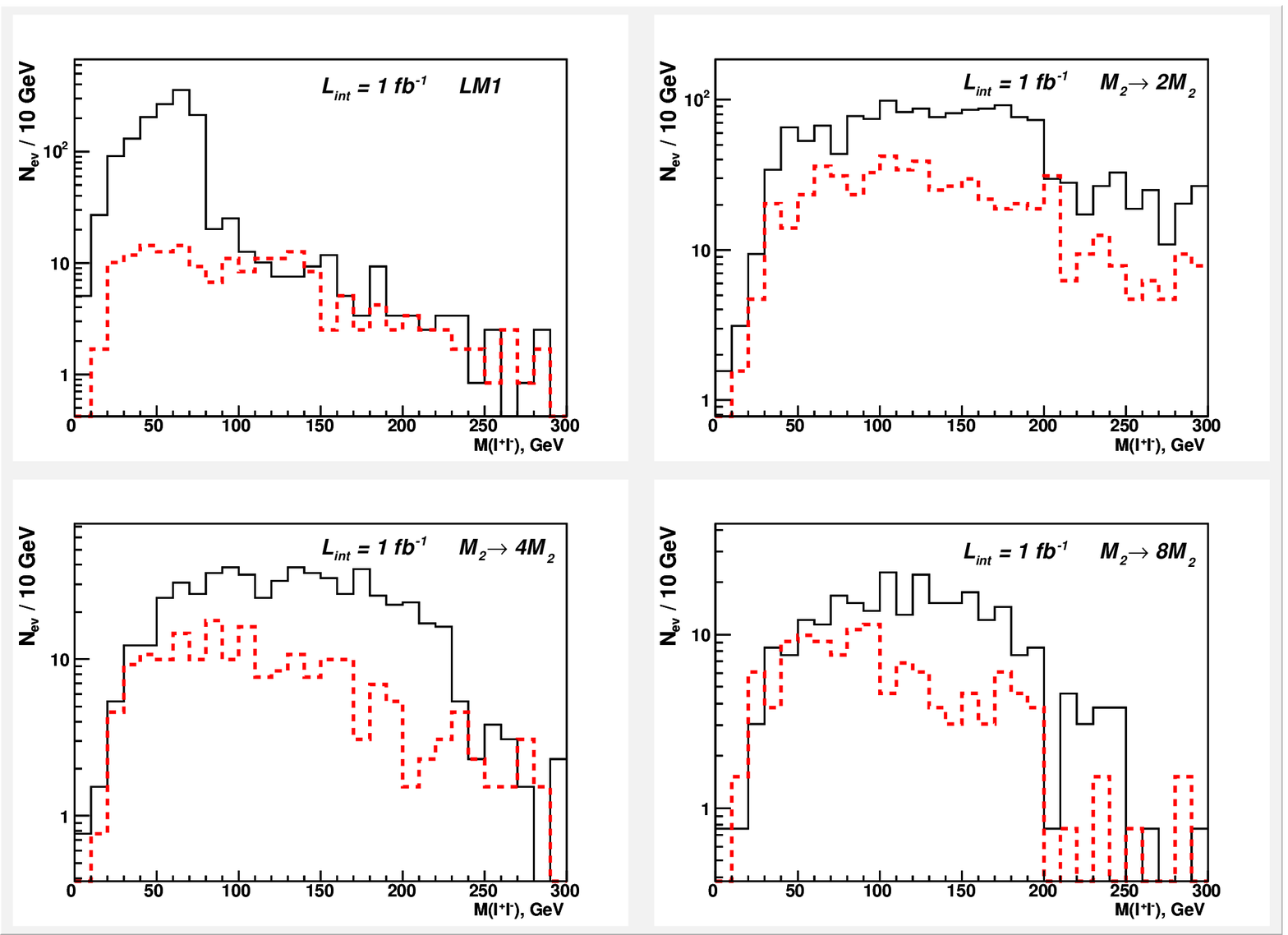}}
    \caption{ The invariant mass distributions of $l^+l^-$  
      (solid line) and $e^{\pm}\mu^{\mp}$ (dotted line) lepton pairs 
      at the extensions of the   point LM1. }
    \label{pM2_MinvBSM4} 
  \end{center}
\end{figure}

Points  (1 - 3) differ from CMS mSUGRA test point LM1 by $SU(2)$ 
gaugino mass $M_2$. The dependence of $m_{inv}(l^+l^-)$ and 
$m_{inv}(e^{\pm}\mu^{\mp})$ invariant spectra 
for the point LM1 and the points  (1 - 3) are shown in Fig.\ref{pM2_MinvBSM4}.
With the increase of $SU(2)$ gaugino mass $M_2$ we see the sharp 
change in the dilepton invariant spectra. 
Points (4 - 6) differ from  CMS mSUGRA test point LM1 by both $SU(2)$ 
gaugino mass and higgsino mass parameter $\mu$.
For the points (7 - 10) the difference with point LM1 also includes the 
increase in gluino and LSP masses.

For the points (7 - 12) we have very sharp change in the number of 
dilepton events. For the luminosity ${\cal L} = 1~fb^{-1}$ after the cuts 
with $p_T^{lept} >20~GeV$, $E_T^{miss} >300~GeV$ the number of events 
without isolated leptons, with single isolated lepton and with two 
isolated leptons are presented in Table \ref{Leptons}. 

\begin{table}[!htb]
\begin{center}
\caption{ The number of  events with  $l = 0$, $l =1$, $l = 2$ 
for background and points LM1, 1-12.}
\begin{tabular}{cccc}
\hline
\hline
~Point~~~& $N(l=0)$& $N(l = 1)$ & $N(l =2)$       \\
\hline
Back.& 20 & 524 & 652 \\
LM1 & 2119& 1278 & 166 \\
1 & 1856 & 1950 & 430  \\
2 & 1858 & 1104 & 187  \\
3 & 1800 & 671 & 83  \\
4 & 1910 & 1876 & 407   \\
5 & 2002 & 513 & 34   \\
6 & 1950 & 449 & 27 \\
7 & 143 & 35 & 1   \\
8 & 87 & 21 & 0 \\
9 & 95 & 21 & 1  \\
10 & 56   & 6 & 0      \\
11 & 1587  &  11 & 1   \\
12 & 1430   & 10 & 1   \\
\hline
\hline
\end{tabular}
\label{Leptons}
\end{center}
\end{table}

As it follows from Table \ref{Leptons} for points (7 - 12) the 
perspective to 
discover SUSY using the signature with single lepton or dileptons 
plus $E^{miss}_T$ looks hopeless. Moreover  for the points (7 - 12) 
the use of more powerful from the SUSY discovery point of view 
signature $no~leptons~ + ~jets~ + ~E_{T}^{miss}$ also looks very 
problematic. 

\begin{figure}[hbtp]
  \begin{center}
    \resizebox{13cm}{!}{\includegraphics{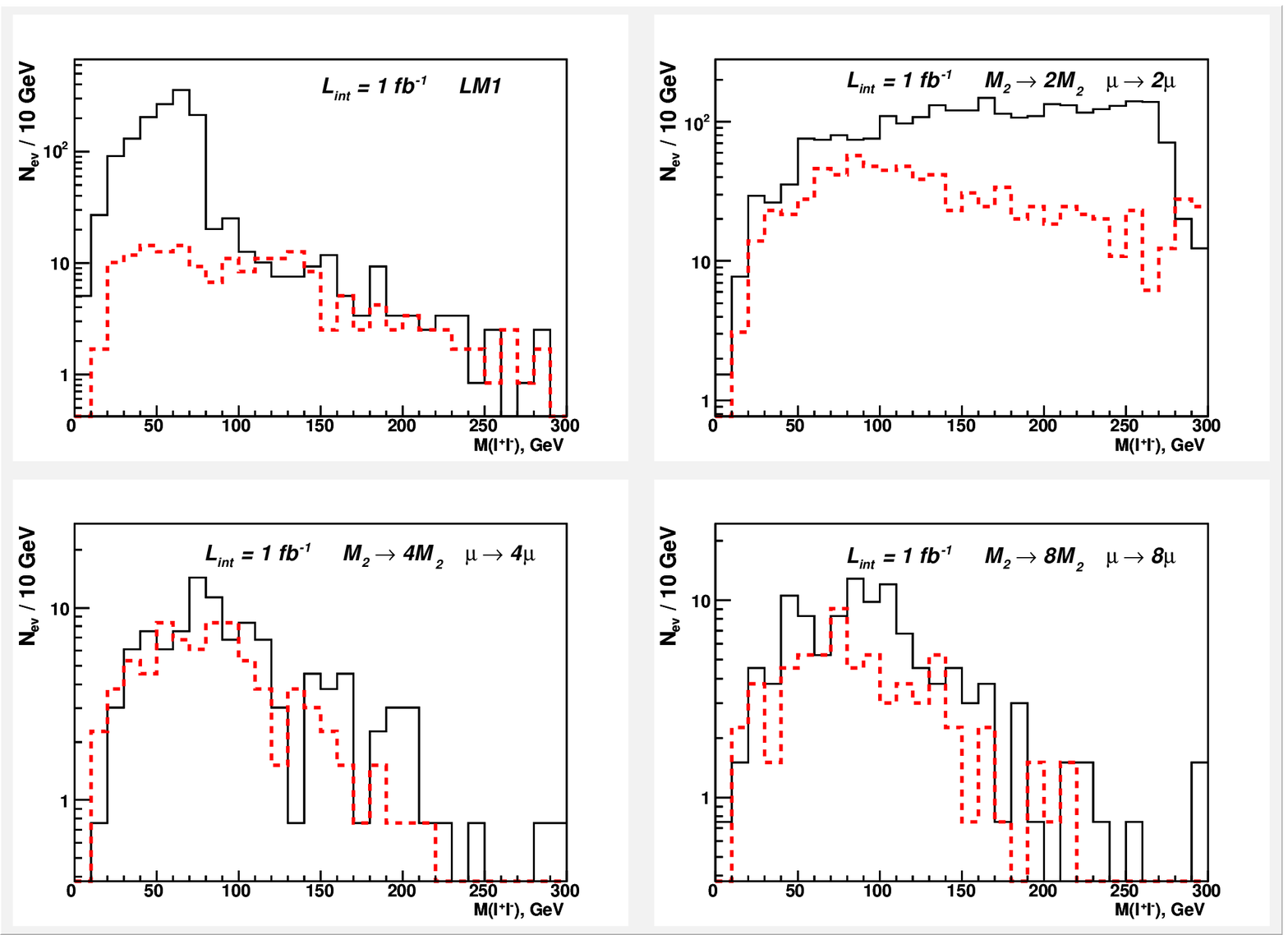}}
    \caption{ The invariant mass distributions of $l^+l^-$   
      (solid line) and $e^{\pm}\mu^{\mp}$ (dotted line) lepton pairs 
      at the extensions of the point LM1. }
    \label{pM2mu_MinvBSM4} 
  \end{center}
\end{figure}

\begin{figure}[hbtp]
  \begin{center}
    \resizebox{13cm}{!}{\includegraphics{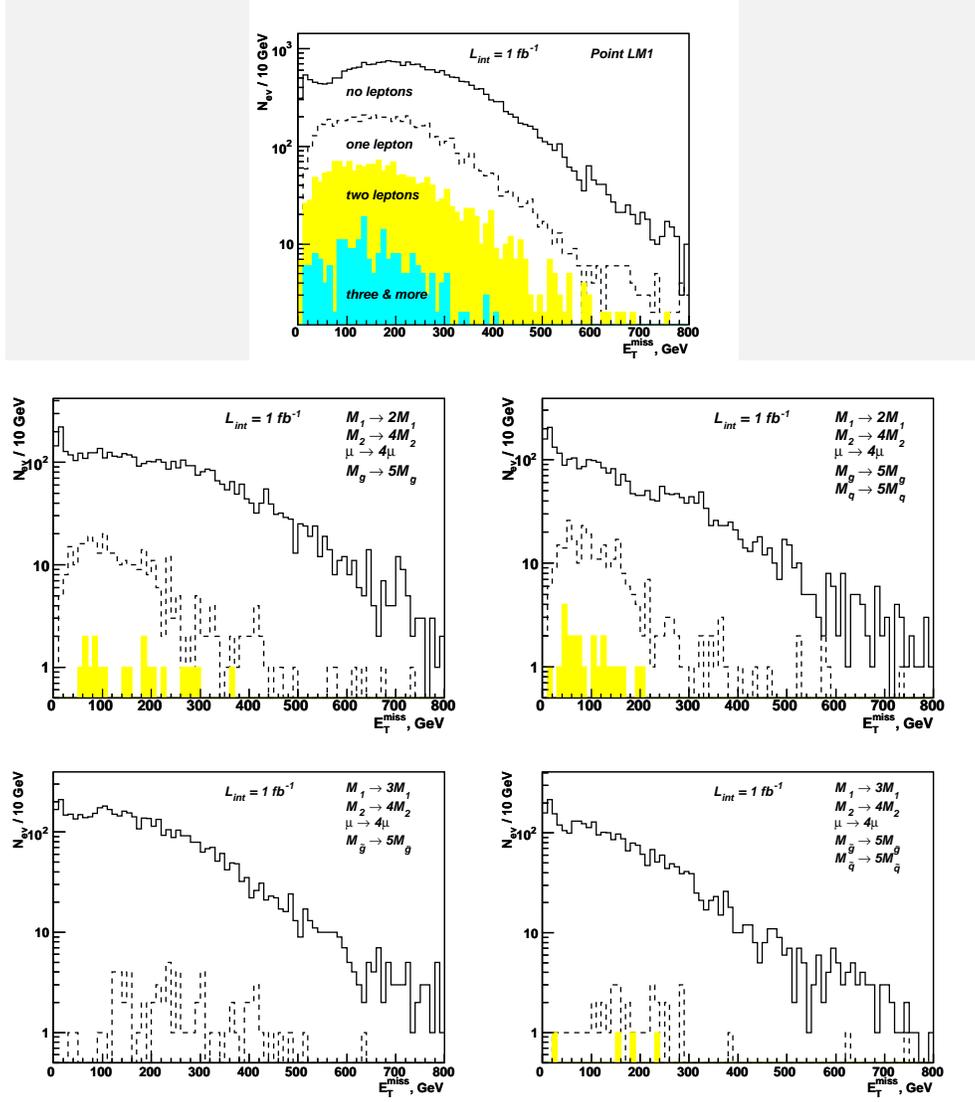}}
    \caption{ The $E^{miss*}_T$ distributions for different 
      modifications of the point LM1. }
    \label{pPx4425} 
  \end{center}
\end{figure}

\begin{figure}[hbtp]
  \begin{center}
    \resizebox{13cm}{!}{\includegraphics{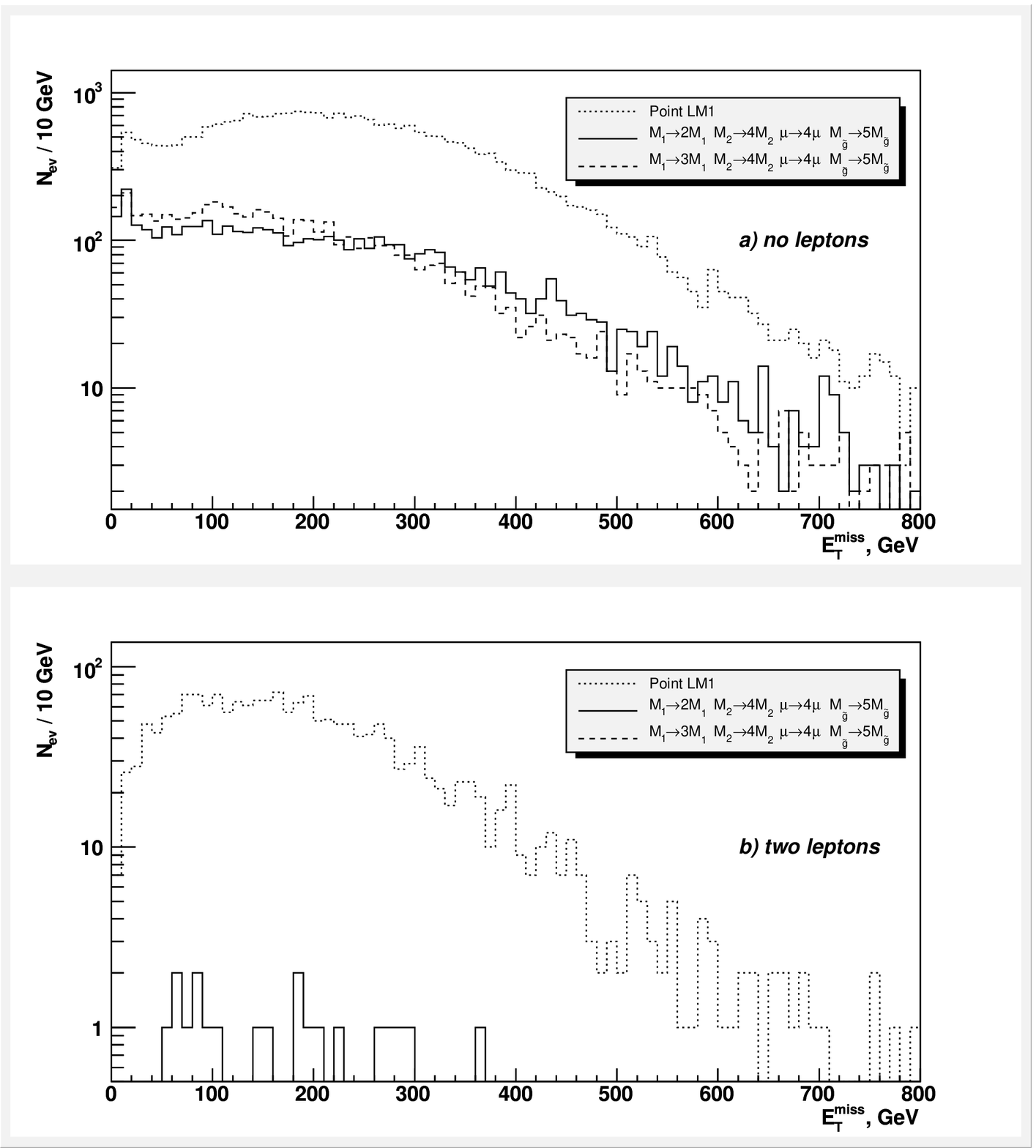}}
    \caption{ The $E^{miss}_T$ distributions for the points LM1, 7 and 8
       for the cases of no leptons and two leptons with $p_T^{lept}>20~GeV$. }
    \label{pL02x4425} 
  \end{center}
\end{figure}

\begin{figure}[hbtp]
  \begin{center}
    \resizebox{10cm}{!}{\includegraphics{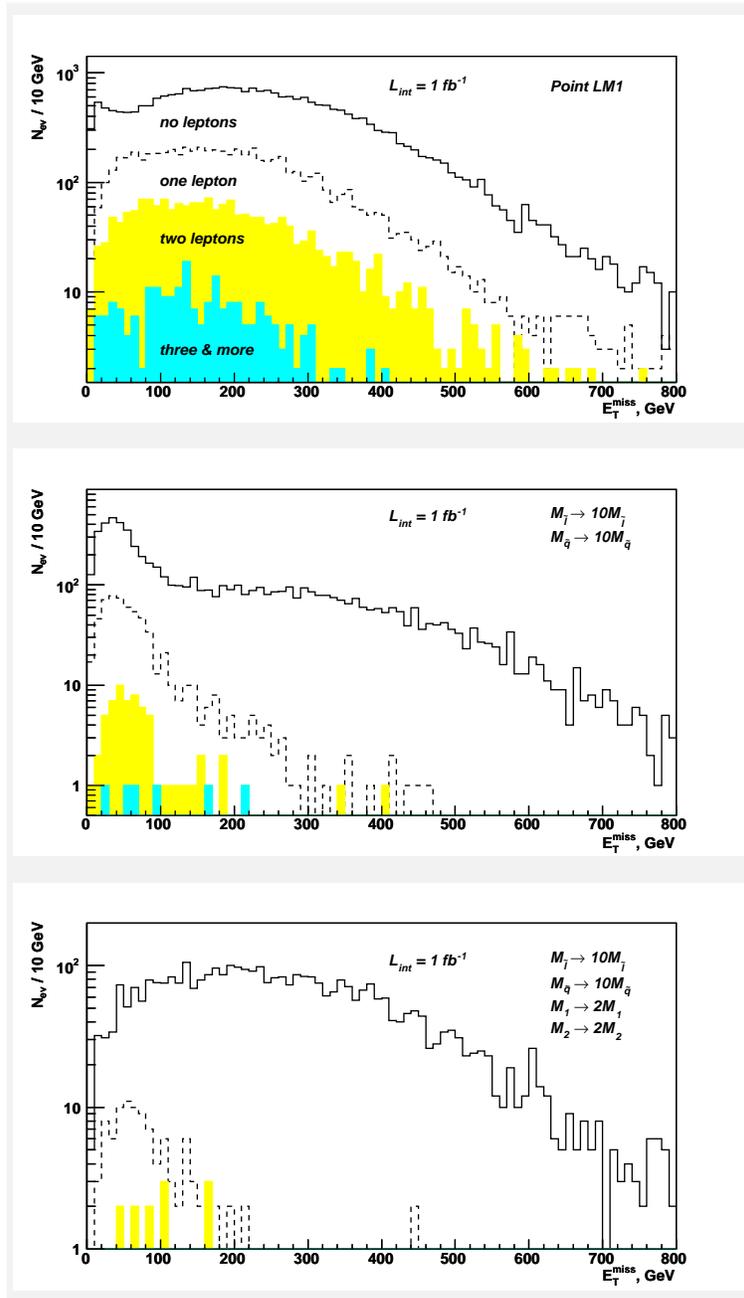}}
    \caption{ The $E^{miss}_T$  distributions for 
      the points LM1, 11 and 12. }
    \label{pPx102} 
  \end{center}
\end{figure}

\begin{figure}[hbtp]
  \begin{center}
    \resizebox{13cm}{!}{\includegraphics{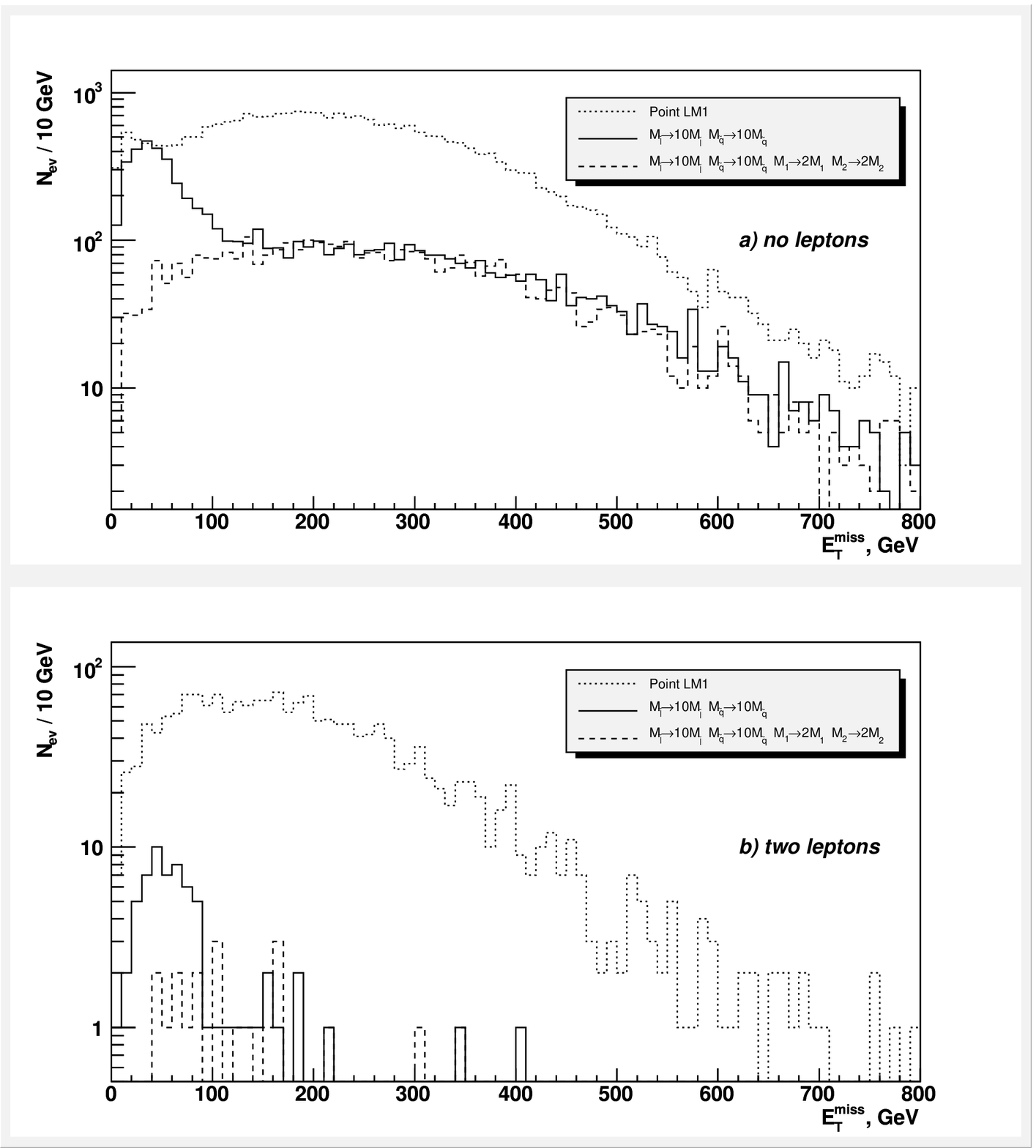}}
    \caption{ The $E^{miss}_T$ distributions  for points LM1, 11 and 12 
      for the cases of no leptons and two leptons with $p_T^{lept} > 20~GeV$. }
    \label{pL02x102} 
  \end{center}
\end{figure}

The invariant mass distributions of $l^+l^-$ (solid line) 
and $e^{\pm}\mu^{\mp}$ (dotted line) 
lepton pairs at the extensions of the  point LM1 are shown in 
Figs.\ref{pM2_MinvBSM4}-\ref{pM2mu_MinvBSM4}.
The distributions of $0$ leptons, $1$ lepton, $2$ leptons and  
$\geq 3$ leptons on $E^{miss}_T$ for the modifications of the 
point LM1  are shown in Figs.\ref{pPx4425} - \ref{pL02x102} for 
an integral luminosity ${\cal L} = 1~fb^{-1}$. 
As one can see from Fig.\ref{pPx4425} for the points (7-10) 
there is sharp change in lepton spectra distributions 
on $E^{miss}_T$ compared to the test point LM1, namely, the number 
of lepton events for the points (7 - 10) are much smaller than for 
the test point LM1. The same situation takes place for the 
$E^{miss}_T$ spectrum of events without leptons, see Fig.\ref{pL02x4425}.

\section{
\label{Conc}
                        Conclusion
}

In this paper we  studied the  $e^{\pm}\mu^{\mp} ~+~E^{miss}_T$
and $l^+l^- ~+~E^{miss}_T$ signatures $(l = e, ~\mu)$ for different parameters 
$\tan\beta$ of the MSSM model. With $\tan\beta$ rising, we observed a 
characteristic change in the shape of dilepton mass spectra in 
$l^+l^- + E^{miss}_T $ versus $e^{\pm}\mu^{\mp}$ final states 
reflecting the decrease of 
$\tilde{\chi}^0_2 \rightarrow l^+l^- \tilde{\chi^0_1}$ branching ratio.
We also studied some non mSUGRA modifications of the point LM1. We have found  the 
drastic change in dilepton spectra with the increase of gaugino masses.  
For the points (7 - 12) the perspective to discover SUSY using the signatures 
with leptons and $E^{miss}_T$ looks hopeless.
Even more powerful signature with $no~ leptons~ + ~ jets ~+~ E^{miss}_T$ looks 
rather pessimistic.  

This work was supported by RFFI grant No 07-02-00256.

\newpage

\end{document}